\documentclass[a4paper,fleqn]{cas-dc}

\usepackage[numbers,sort&compress]{natbib}
\usepackage[utf8]{inputenc}
\usepackage{amsmath}
\usepackage{mathrsfs}
\usepackage{amssymb}
\usepackage{color}
\usepackage{dsfont}
\usepackage{comment}
\usepackage{graphicx}
\usepackage[font = footnotesize]{caption}
\usepackage[font = footnotesize]{subcaption}
\usepackage{mathStyle}
\usepackage{color}
\usepackage{tikz}
\usepackage{booktabs}
\usepackage{tabularx}
\usepackage{multirow}
\usepackage{bm}
\usepackage{makecell}
\usepackage{indentfirst}
\usepackage{booktabs}

\usepackage{color, colortbl}
\definecolor{Yellow}{rgb}{1,1,0}

\usepackage{hhline}

\usetikzlibrary{shapes,arrows,calc,positioning,patterns}
\tikzstyle{bigblock} = [draw, fill=blue!20, rectangle, 
    minimum height=6em, minimum width=8em]
\tikzstyle{medblock} = [draw, fill=blue!20, rectangle, 
    minimum height=4em, minimum width=4em]    
\tikzstyle{mux} = [draw, fill=black!20, rectangle, 
    minimum height=5em, minimum width=0.1em]    
\tikzstyle{smallblock} = [draw, fill=blue!20, rectangle, 
    minimum height=3em, minimum width=4em]
\tikzstyle{sum} = [draw, fill=blue!20, circle, node distance=1cm]
\tikzstyle{signal} = [coordinate]
\tikzstyle{pinstyle} = [pin edge={to-,thin,black}]
\tikzstyle{block} = [draw, fill=blue!20, rectangle, 
    minimum height=3em, minimum width=6em]
\tikzstyle{blockS} = [draw, fill=blue!20, rectangle, 
    minimum height=3em, minimum width=4em]  
\tikzstyle{sum} = [draw, fill=blue!20, circle, node distance=1.5cm]
\tikzstyle{gain} = [draw, fill=blue!20, regular polygon, regular polygon sides = 3, node distance=1.25cm, shape border rotate = -90]
\tikzstyle{mult} = [draw, fill=blue!20, circle, node distance=1.25cm ,inner sep=0pt, minimum size = 0.3cm]

\tikzstyle{input} = [coordinate]
\tikzstyle{output} = [coordinate]
\tikzstyle{spring}=[thick,decorate,decoration={zigzag,pre length=0.3cm,post length=0.3cm,segment length=6}]
\tikzstyle{damper}=[thick,decoration={markings,  
  mark connection node=dmp,
  mark=at position 0.5 with 
  {
    \node (dmp) [thick,inner sep=0pt,transform shape,rotate=-90,minimum width=15pt,minimum height=3pt,draw=none] {};
    \draw [thick] ($(dmp.north east)+(2pt,0)$) -- (dmp.south east) -- (dmp.south west) -- ($(dmp.north west)+(2pt,0)$);
    \draw [thick] ($(dmp.north)+(0,-5pt)$) -- ($(dmp.north)+(0,5pt)$);
  }
}, decorate]
\tikzstyle{ground}=[fill,pattern=north east lines,draw=none,minimum width=0.75cm,minimum height=0.3cm]

\usepackage[linesnumbered,ruled,vlined]{algorithm2e}
\SetKwInOut{KwParam}{Parameters}
\let\oldnl\nl
\newcommand{\nonl}{\renewcommand{\nl}{\let\nl\oldnl}}

\newcounter{example}

\usepackage{hyperref}
\usepackage{xcolor}
\hypersetup{
    colorlinks,
    linkcolor={blue!100!black},
    citecolor={blue!50!black},
    urlcolor={blue!80!black}
}

\begin{document}
\let\WriteBookmarks\relax
\def\floatpagepagefraction{1}
\def\textpagefraction{.001}
\shorttitle{Performance-Based Adaptation Termination for Mitigating Parameter Drift in Adaptive Control}
\shortauthors{J.A. Paredes Salazar et~al.}

\title[mode = title]{Performance-based Adaptation Termination for Preventing \\ Parameter Drift in Adaptive Vibration Suppression}

\author[1]{Juan Augusto {Paredes Salazar}}[orcid=0000-0001-7486-1231]
\cormark[1]
\ead{japarede@umbc.edu}
\credit{Conceptualization, Methodology, Software, Writing - Original draft preparation, Writing – review and editing}

\author[1]{Ankit Goel}[orcid=0000-0002-4146-6275]
\ead{ankgoel@umbc.edu}
\credit{Funding acquisition, Supervision, Writing - Original draft preparation, Writing – review and editing}

\affiliation[1]{organization={Department of Mechanical Engineering, University of Maryland Baltimore County},
                addressline={1000 Hilltop Circle}, 
                city={Baltimore},
                postcode={21250}, 
                state={Maryland},
                country={United States}}

\cortext[cor1]{Corresponding author}

\begin{abstract}
Parameter drift remains a practical limitation of adaptive vibration control systems, particularly when persistence of excitation diminishes after disturbance attenuation or when measurement disturbances render the adaptation problem ill-conditioned. 
This issue is especially relevant in flexible structures, where adaptive controllers may continue updating parameters even after satisfactory vibration suppression has been achieved. 
This paper proposes a computationally efficient root-mean-square (RMS)-based stopping criterion that mitigates parameter drift by freezing adaptation once satisfactory vibration attenuation has been sustained over a prescribed interval. 
The criterion is implemented by recursively computing an exponentially weighted moving RMS of the performance variable and requires negligible additional computational effort. 
The proposed mechanism is integrated with retrospective cost adaptive control (RCAC) and validated through both numerical simulations and closed-loop experiments on a cantilever-beam vibration-suppression platform with a noncollocated actuator configuration.
Without the proposed stopping criterion, continued adaptation after disturbance rejection leads to gradual parameter drift and degradation of vibration suppression performance. 
When the RMS-based threshold is enabled, controller parameters remain bounded, and the achieved vibration attenuation is preserved. 
The results demonstrate that a simple performance-based monitoring mechanism can effectively prevent parameter drift in adaptive vibration control, retain the transient benefits of adaptation, and incur minimal computational overhead.
\end{abstract}


\maketitle

\section{Introduction} \label{sec:introduction}

Adaptive control has been widely investigated for vibration suppression in flexible structures subject to uncertain or poorly modeled dynamics.
In many vibration-control applications, including flexible beams, plates, and aerospace structures, disturbances may excite lightly damped structural modes whose dynamics are difficult to model accurately \cite{balas1978active, meirovitch1991dynamics, preumont1997vibration, balas2003feedback, sumer2012adaptive}.
These resonant responses can lead to sustained oscillations, fatigue loading, and degraded performance if not effectively mitigated \cite{balas1978active, meirovitch1991dynamics, preumont1997vibration, gawronski2004, song2006, giurgiutiu2008, takacs2011model, inman2014}. 
Adaptive control algorithms provide an attractive alternative to model-based approaches because they can automatically tune controller parameters to attenuate sustained disturbances using measured data rather than an explicit plant model.

However, a well-known practical limitation of adaptive control is \emph{parameter drift}.
After satisfactory vibration attenuation is achieved, the effective excitation perceived by the adaptive law may decrease significantly.
In vibration suppression problems, the oscillatory response may become small once vibration attenuation  is achieved, thereby reducing the information available to the adaptation mechanism.
In the presence of measurement noise or bounded disturbances of sufficient amplitude, the underlying optimization problem can become ill-conditioned, and the parameters updated by the adaptation law may continue to evolve unnecessarily or even diverge, potentially yielding unstable closed-loop behavior.
This phenomenon is referred to as parameter drift \citep{egardt1979,anderson1985,ioannou1996,lawrence2002parameter}.

The parameter drift problem is particularly challenging in adaptive vibration suppression approaches because persistent oscillatory responses can exacerbate it \citep{landau1999robust, karimi2002robust, slotine2002adaptive, zhang2005hybrid, manosa2005control, chang2007low, trajkov2008direct, ma2009adaptive, coza2011adaptive, coppola2015experimental, macnab2016, jiawei2019robust, zhang2023optimal, zhang2025adaptive, herburger2025adaptive, huo2025finite, jiang2025review, piramoon2025design}.
In the context of classical adaptive control, several modifications have been developed to mitigate parameter drift.
Table \ref{tab:adaptive_drift_methods} summarizes several commonly used approaches and contrasts them with the proposed approach.  
While these approaches improve robustness to disturbances and noise, they alter the nominal adaptive structure and typically introduce additional tuning parameters. 
As a result, they entail a tradeoff between robustness to measurement disturbances and disturbance-rejection performance.

\begin{table}[h]
\centering
\footnotesize
\setlength{\tabcolsep}{1pt}
\caption{Common modifications used to mitigate parameter drift in adaptive control.}
\label{tab:adaptive_drift_methods}
\begin{tabular}{p{0.20\columnwidth} p{0.18\columnwidth} p{0.6\columnwidth}}
\toprule
\textbf{Approach} & \textbf{References} & \textbf{Description} \\
\midrule

Normalization signals 
&
\citep{praly1984robust,ioannou2003robust,kreisselmeier2003robust}
&
Scale the adaptation law using regressor-based normalization to limit large parameter updates.

\textit{However, these methods modify the adaptive update law and require careful tuning of normalization signals, which may affect adaptation performance. }
\\ \cmidrule(lr){1-3}
Projection operators 
&
\citep{goodwin1987parameter,naik1992robust,yao1997high}
&
Restrict parameter estimates to a bounded set to prevent drift. 

\textit{However, projection modifies the parameter-update mechanism and requires prior specification of parameter bounds, which may be difficult to determine in practice. }
\\ \cmidrule(lr){1-3}

Deadzone nonlinearities 
&
\citep{ydstie1989,xu2000adaptive,charandabi2011improved,bagherpoor2015robust}
&
Suspend adaptation when the tracking error is below a specified threshold. 

\textit{However, the deadzone is a static system that reacts to the instantaneous value of the error, which is not suited to oscillatory signals. Furthermore, the deadzone may prematurely halt adaptation or degrade disturbance rejection. }
\\ \cmidrule(lr){1-3}

$e$-modification 
&
\citep{yucelen2010kalman,hoang2016neural,xie2025variable}
&
Add a leakage term proportional to the parameter estimate.

\textit{However, the leakage term alters the nominal adaptive law and may introduce steady-state bias in the parameter estimates. }
\\ \cmidrule(lr){1-3}

$\sigma$-modification 
&
\citep{ioannou1983,ioannou1986,tsakalis1992sigma,he2013asymptotic}
&
Introduce damping in the adaptation law to reduce parameter growth. 

\textit{However, the additional damping modifies the adaptation dynamics and can slow parameter convergence. }
\\ \cmidrule(lr){1-3}

Controller-output filtering 
&
\citep{kharisov2010,hovakimyan2010}
&
Filter control signals to reduce disturbance effects on adaptation. 

\textit{However, filtering introduces additional dynamics and may reduce the responsiveness of the closed-loop system. }
\\ \cmidrule(lr){1-3}

Controller-output averaging 
&
\citep{macnab2016,nicol2011}
&
Use averaged control signals to reduce oscillatory disturbances.

\textit{However, averaging relies on time-window selection and may obscure transient behavior important for adaptation. }
\\ 

\midrule

\textbf{This work}
&
\multicolumn{2}{p{0.78\columnwidth}}{
\textbf{Mitigates parameter drift without altering the adaptive update law. 
Instead, a performance-based stopping mechanism terminates adaptation when an exponentially weighted RMS measure of the vibration response remains below a prescribed threshold for a sustained interval, after which the controller operates with fixed gains.}
} \\

\bottomrule
\end{tabular}
\end{table}

This paper proposes a performance-based mechanism to mitigate parameter drift without modifying the adaptive update law itself.
The key idea is to terminate adaptation once satisfactory vibration attenuation  has been sustained over a \textit{prescribed interval}. 
To accomplish this, we introduce a root-mean-square (RMS)-based stopping criterion computed recursively from the measured performance variable.
The use of an RMS metric is particularly well-suited for vibration suppression problems because structural vibration responses are inherently oscillatory and typically do not converge to zero even under effective vibration attenuation .
Instead, the RMS value provides a direct measure of the energy of the oscillatory response, making it a natural indicator of sustained vibration attenuation.

When the exponentially weighted RMS falls below a user-defined threshold, parameter updates are frozen, and the controller transitions from adaptive to fixed-gain operation.
The proposed mechanism is computationally lightweight, requires minimal tuning, and relies only on measured data already available in the control loop.
Note that variations of the RMS-based criteria have been widely used for fault detection and health monitoring \citep{ding2008model,abdo2011robust,khan2011threshold,donnelly2015rms,igba2016analysing,hong2016vibration,liu2021model,mohd2021vibration}, demonstrating their suitability as a practical performance measure.

In this work, the stopping criterion is integrated with Retrospective Cost Adaptive Control (RCAC) and evaluated for vibration suppression of a cantilever beam subject to harmonic disturbance.
Both numerical simulations and closed-loop experiments with a noncollocated actuator configuration are presented.
Without the proposed stopping criterion, continued adaptation after disturbance attenuation leads to gradual parameter drift and performance degradation.
With the RMS-based mechanism enabled, controller parameters remain bounded and the achieved disturbance-rejection performance is preserved.

The results demonstrate that performance-based monitoring can effectively prevent parameter drift while retaining the benefits of adaptive control in real-time applications.
The proposed approach does not require an explicit plant model, does not increase computational complexity, and can be incorporated into a broad class of adaptive controllers.
The key contributions of this work are summarized as follows.
\begin{enumerate}

\item \textbf{Performance-based adaptation termination for mitigating parameter drift.}
A performance-governed mechanism is proposed to mitigate parameter drift in adaptive vibration control systems without modifying the nominal adaptive update law. 
Instead of altering the parameter-update dynamics through projection operators, normalization schemes, or leakage modifications, the proposed approach regulates \textit{when adaptation is permitted to continue} by monitoring measured closed-loop performance.

\item \textbf{Recursive RMS-based performance metric for vibration suppression.}
A computationally efficient root-mean-square (RMS) performance metric is developed based on an exponentially weighted moving average. 
The resulting formulation admits a recursive implementation and an interpretation in terms of an equivalent moving-window length, which provides intuitive tuning guidelines and prevents premature adaptation termination due to transient noise. 
Because vibration responses are inherently oscillatory, the RMS metric provides a natural measure of sustained vibration attenuation.

\item \textbf{Simulation and experimental validation on a flexible structure.}
The proposed stopping criterion is integrated with Retrospective Cost Adaptive Control (RCAC) and validated through both numerical simulations and real-time experiments on a cantilever beam with a noncollocated actuator configuration. 
The results demonstrate that continued adaptation after vibration attenuation  can lead to parameter drift and degraded vibration suppression, whereas the proposed RMS-based mechanism prevents destabilizing parameter evolution while preserving the achieved closed-loop performance.

\end{enumerate}

The remainder of this paper is organized as follows.
Section \ref{sec:prob} formulates the sampled-data vibration attenuation  problem for a continuous-time system.
Section \ref{sec:RCAC} summarizes the Retrospective Cost Adaptive Control (RCAC) framework used in this work.
Section \ref{sec:RMS_freeze} introduces the proposed root-mean-square (RMS)-based learning stopping criterion for mitigating parameter drift.
Section \ref{sec:simulation_experiment} presents the numerical simulation model and experimental setup, along with closed-loop simulation and experimental results.
Finally, Section \ref{sec:conclusions} concludes the paper.

\section{Sampled-Data Disturbance Rejection Problem}
\label{sec:prob}

This paper focuses on vibration suppression for a flexible structure under sampled-data control. In particular, we consider the suppression of externally induced vibrations in a continuous-time system $\SM$ subject to a bounded disturbance $d(t)$.

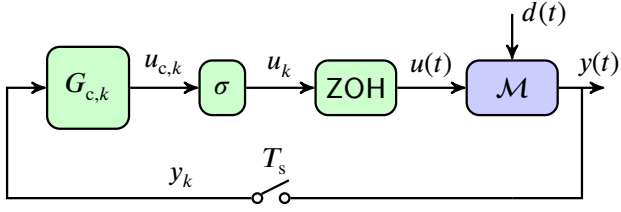
\begin{figure}
    \centering
    \vspace{-0.25em}
    \resizebox{\columnwidth}{!}{%
    \begin{tikzpicture}[>={stealth'}, line width = 0.25mm]
    \node [input, name=ref]{};
    \node [smallblock, fill=green!20, rounded corners, right = 0.5cm of ref , minimum height = 0.9cm, minimum width = 0.9cm] (controller) {$G_{\rmc, k}$};
    %
    %
    %
    \node[smallblock, fill=green!20, rounded corners, right = 0.75 of controller, minimum height = 0.6cm, minimum width = 0.5cm](sat_Blk){$\sigma$};
    \node [smallblock, fill=green!20, rounded corners, right = 0.75cm of sat_Blk, minimum height = 0.6cm , minimum width = 0.5cm] (DA) {ZOH};
    
    \node [smallblock, rounded corners, right = 0.75cm of DA, minimum height = 0.6cm , minimum width = 1cm] (system) {$\SM$};
    \node [output, right = 0.5cm of system] (output) {};
    \node [input, below = 0.9cm of system] (midpoint) {};
    
    \draw [->] (controller) -- node [above] {\small $u_{\rmc,k}$} (sat_Blk);
    \draw [->] (sat_Blk) -- node [above] {\small $u_k$} (DA);\
    \draw [->] (DA) -- node [above] {\small $u (t)$} (system);
    
    \node[circle,draw=black, fill=white, inner sep=0pt,minimum size=3pt] (rc11) at ([xshift=-2.5cm]midpoint) {};
    \node[circle,draw=black, fill=white, inner sep=0pt,minimum size=3pt] (rc21) at ([xshift=-2.8cm]midpoint) {};
    \draw [-] (rc21.north east) --node[below,yshift=.55cm]{$T_\rms$} ([xshift=.3cm,yshift=.15cm]rc21.north east) {};
    
    \draw [->] (system) -- node [name=y, near end]{} node [very near end, above] {\small $y (t)$}(output);

    \draw [->] ([yshift = 0.5cm]system.north) -- node [xshift = 0.35cm,yshift = 0.25 cm] {\small $d(t)$} (system.north);
    
    \draw [-] (y.west) |- (midpoint);
    \draw [-] (midpoint) -| node [above, xshift=-1.2cm] {\small $y_k$} (rc11.east);
    \draw [->] (rc21) -- +(-2.75,0)|- (controller.west) ;
    
    \end{tikzpicture}
    }  
    \caption{Sampled-data disturbance-rejection architecture for the continuous-time system $\SM$ with control input $u$, external disturbance $d$, measured output $y$, and actuator constraints $\sigma.$
    %
    %
    }
    \label{fig:RT_CT_blk_diag}
    \vspace{-0.5em}
\end{figure}

The control architecture is shown in Figure \ref{fig:RT_CT_blk_diag}.
The plant $\SM$ has input $u(t) \in \BBR$, disturbance input $d(t) \in \BBR$, and measured output $y(t) \in \BBR$. The output is sampled with sampling period $T_\rms > 0$, yielding $y_k \isdef y(kT_\rms).$
The objective is disturbance attenuation, that is, to reduce the amplitude of $y(t)$ despite sustained external excitation.

The sampled measurement $y_k$ is used to define the performance variable $z_k$, which is described in the next section and is used by the discrete-time controller $G_{\rmc,k}$ to compute the commanded control $u_{\rmc,k}$.
Because real actuators are subject to magnitude and rate constraints, the implemented control input is 
\begin{equation}
    u_k \isdef \sigma(u_{\rmc,k}),
\end{equation}
where $\sigma(\cdot)$ denotes actuator saturation.
The continuous-time control signal is obtained through a zero-order hold, that is, for $t \in [kT_\rms,(k+1)T_\rms),$
    $u(t) = u_k.$

The controller $G_{\rmc,k}$ is adaptive and updates its parameters online based on measured data.
Under sustained disturbance excitation, the controller initially adapts to reduce vibration amplitude.
However, once attenuation is achieved, the effective excitation seen by the adaptive law may decrease significantly.
In the presence of measurement noise or unmodeled dynamics, this loss of excitation can render the underlying parameter update problem ill-conditioned, leading to parameter drift.
The central problem addressed in this paper is how to automatically terminate adaptation once satisfactory vibration suppression  has been achieved, to prevent parameter drift while preserving the achieved closed-loop performance and stability.

\section{Retrospective Cost Adaptive Control for Disturbance Rejection}
\label{sec:RCAC}
Retrospective Cost Adaptive Control (RCAC) is described in detail in \citep{rahmanCSM2017}. 
Here we summarize the elements relevant to disturbance rejection and to the stopping mechanism proposed in Section \ref{sec:RMS_freeze}.

\subsection{Controller}
Consider the strictly proper, discrete-time, input-output controller  
\begin{align}
    u_{\rmc, k} = \sum_{i=1}^{l_\rmc}P_{i,k}u_{k-i} + \sum_{i=1}^{l_\rmc}Q_{i,k}z_{k-i}, \label{IO_controller}
\end{align}
where $u_{\rmc, k} \in \mathbb R^{l_u}$ is the commanded input and the controller output $u_k \in \mathbb R^{l_u}$ is the control input,  $z_k \in \mathbb R^{l_z}$ is the measured performance variable, $l_\rmc$ is the controller-window length, and, for all $i\in \{1,\ldots,l_\rmc\},$  $P_{i,k} \in \mathbb R^{l_u \times l_u}$ and $Q_{i,k} \in \mathbb R^{l_u \times l_z}$ are the controller coefficient matrices.
Note that  $u_k$ results from applying constraints to $u_{\rmc, k},$ as shown in Section \ref{sec:prob}.
The controller \eqref{IO_controller} can be written as
\begin{align}
    u_{\rmc, k}   =   \phi_k  \theta_k , \label{eq:controller}
\end{align}
where 
\begin{align}
	\phi_k &\isdef
    \left[ \arraycolsep=3pt\def\arraystretch{0.9} \begin{array}{cccccc} 
    			u_{k-1}^\rmT & \cdots & u_{k-l_\rmc}^\rmT & z_{k-1}^\rmT & \cdots & z_{k-l_\rmc}^\rmT
    \end{array} \right]
    		\otimes
    		I_{l_u}
    		\nn \\
            &\in \mathbb{R}^{l_u \times l_{\theta}},  \label{eq:phik} \\
    	\theta_k &\isdef {\rm vec}
    \left[ \arraycolsep=1.7pt\def\arraystretch{0.9} \begin{array}{cccccc} 
        P_{1,k} &\cdots &P_{l_\rmc,k} &Q_{1,k} &\cdots &Q_{l_\rmc,k}
    \end{array} \right] \in \BBR^{l_\theta},
\end{align}
$l_\theta \isdef l_\rmc l_u (l_u + l_z),$ and $\theta_k$ is the vector of controller coefficients, which are updated at each time step $k$.
Note that 
$\vek X\in\BBR^{nm}$ denotes the vector formed by stacking the columns of $X\in\BBR^{n\times m}$, and 
$\otimes$ denotes the Kronecker product.

The controller coefficients $\theta_k$ are optimized using the retrospective cost described below to ensure asymptotic stability of the closed-loop system.
In the context of the disturbance rejection problem, the performance variable is defined as $z_k \isdef K y_k,$ where $K$ is a constant gain.
In this work, $K$ is selected so that $z_k$ is expressed in millimeters. 

\subsection{Retrospective Performance}
Next, define the retrospective cost variable
\begin{align}
	\hat z_k (\hat \theta) \isdef z_k  - G_\rmf(\textbf{q})(u_k - \phi_k \hat{\theta}), \label{zhat1}
\end{align}
where $G_\rmf(\bfq)$ is an asymptotically stable, strictly proper transfer function and $\hat \theta \in \mathbb{R}^{l_\theta}$ is a candidate controller parameter vector.
Note that $\bfq$ is the forward-shift operator, such that $\bfq u_k = u_{k+1}$ and $\bfq\inv u_k = u_{k-1}.$
The role of \eqref{zhat1} is to evaluate how the closed loop would have behaved if the control input had been $\phi_k \hat \theta$ \cite{santillo2010adaptive}. 
Thus, $G_\rmf$ serves as a closed-loop target model for adaptation.

\subsection{Target Model}

The target model has the form
\begin{align}
    G_\rmf(\bfq) = D_\rmf^{-1}(\bfq) N_\rmf(\bfq),
\end{align}
where $D_\rmf$ and $N_\rmf$ are polynomial matrices and $D_\rmf$ has leading coefficient $I$.
When the plant is SISO, the design of $G_\rmf$ requires knowledge of
(i) the sign of the leading numerator coefficient,
(ii) the relative degree, and
(iii) any nonminimum-phase zeros \citep{rahmanCSM2017,islam2021data}.
To avoid the cancellation of unmodeled nonminimum-phase zeros, a control-weighting term is added to the cost function below.

\subsection{Retrospective Cost}

The retrospective cost is defined as
\begin{align}
J_k(\hat \theta)
&=
\sum_{i=0}^{k}
\hat z_i^{\rmT}(\hat \theta)\hat z_i(\hat \theta)
+
(\phi_i \hat \theta)^{\rmT} R_u \phi_i \hat \theta
\nn \\
&\quad +
(\hat \theta - \theta_0)^{\rmT}
P_0^{-1}
(\hat \theta - \theta_0),
\label{eq:Jg}
\end{align}
where $R_u \geq 0$ is a control-weighting matrix and $P_0 > 0$ is a regularization matrix.
The matrix $R_u$ penalizes excessive control effort and prevents cancellation of unmodeled nonminimum-phase zeros, while $P_0^{-1}$ initializes the recursive update.

\subsection{Recursive Controller Optimization}

Minimization of \eqref{eq:Jg} yields a recursive least-squares (RLS) update of the form
\begin{align}
P_k &= P_{k-1}
- P_{k-1} \Psi_k^{\rmT}
\left(
I + \Psi_k P_{k-1} \Psi_k^{\rmT}
\right)^{-1}
\Psi_k P_{k-1}, \label{eq:pk_update} \\
\theta_k &= \theta_{k-1}
- P_k \Psi_k^{\rmT} \bar R
\matl
z_k - (u_{\rmf,k} - \phi_{\rmf,k} \theta_{k-1}) \\
\phi_k \theta_{k-1}
\matr, \label{eq:theta_update}
\end{align}
where
\begin{align}
    \Psi_k 
        &\isdef
            \matl
            \phi_{\rmf,k} \\
            \phi_k
            \matr,
        \\
\bar R 
    &\isdef
        \mathrm{diag}(I_{l_z}, R_u) \in \BBR^{ (l_z + l_u) \times  (l_z + l_u) }.
\end{align}
and
\begin{align}
    \phi_{\rmf, k} \isdef G_{\rmf} (\bfq) \phi_k,  \label{eq:phifk}
    \\
    u_{\rmf, k} \isdef G_{\rmf} (\bfq) u_k. \label{eq:ufk}
\end{align}
For all simulations and experiments, we set $\theta_0=0_{l_\theta\times 1}$ to reflect the absence of a prior controller, and 
$P_0 = p_0 I$, where $p_0 > 0$ determines the initial adaptation rate.

\subsection{Target Model for Sinusoidal Disturbances}

For vibration suppression with sinusoidal disturbances and time delay, the target model is chosen as
\begin{equation}
G_\rmf(\bfq)
=
\frac{N}{\bfq^{d_\rmf}}
\left(
\frac{1}{\bfq^2
- 2 \alpha_\rmf \cos(\omega_\rmf T_s)\bfq
+ \alpha_\rmf^2}
\right),
\end{equation}
where $N \in \{-1,1\}$ encodes the sign of the leading coefficient,
$d_\rmf \ge 0$ represents time delay,
$\omega_\rmf = 2\pi f_\rmf$ corresponds to the disturbance frequency,
and $\alpha_\rmf \in (0,1]$ determines pole placement relative to the unit circle.
%
Values of $\alpha_\rmf$ closer to 1 place the target-model poles nearer the unit circle, resulting in a more lightly damped internal model of the disturbance. 
In practice, values of $\alpha_\rmf$ closer to 1 may increase sensitivity to noise.

\section{Root Mean Square-Based Learning Stopping Criteria} \label{sec:RMS_freeze}

To prevent parameter drift, this paper proposes stopping the adaptive algorithm from learning once the amplitude of the performance variable remains below a chosen threshold for a sustained interval, rather than triggering instantaneously based on a single sample.
For this purpose, the root-mean-square (RMS) of $z_k$ is chosen as the learning stopping criterion.
Let $\gamma \in (0,1)$ be an exponential forgetting factor.
For $k\ge 2$, define the exponentially weighted moving RMS of the sequence $\{z_k\}$ by
\begin{align}
    z_{{\rm rms},k}
    \isdef
    \sqrt{
        (1-\gamma)\sum_{i=1}^{k-1} \gamma^{\,k-1-i} z_i^2
    }.
    \label{eq:z_rms_def}
\end{align}
Note that $z_{{\rm rms}, k},$ defined in \eqref{eq:z_rms_def}, can be computed recursively as
\begin{align}
    z_{{\rm rms}, k}  
        =
            \sqrt
            {
            \gamma z_{{\rm rms}, k-1}^2 + (1-\gamma) z_{k-1}^2
            }.
    \label{eq:z_k_rms_recursive}
\end{align}

    The definition \eqref{eq:z_rms_def} corresponds to an exponentially weighted moving average (EWMA) of $z_k^2$, and thus emphasizes recent samples while exponentially discounting older samples.
To relate $\gamma$ to an equivalent moving-window length, we define the effective sample size $n_{\rm samp}>0$ by
\begin{align}
    n_{\rm samp} \isdef \frac{\gamma}{1-\gamma},
\end{align}
which implies that 
\begin{align}
    \gamma = \frac{n_{\rm samp}}{n_{\rm samp}+1}.
    \label{eq:gamma_nsamp}
\end{align}
Note that with this choice of $\gamma,$ the recursion given by \eqref{eq:z_k_rms_recursive} acts as a moving average of the sequence $\{z_k^2\}$ over the $n_{\rm samp}$ most recent samples. 
Consequently, a larger value of $n_{\rm samp} $ yields more smoothing of $z_{{\rm rms}, k}  $ and requires  $z_k^2$ to remain small for a longer duration for $z_{{\rm rms}, k}  $ to fall below a prescribed threshold value. 
Note that the uniform moving-window RMS over $n_{\rm samp}$ samples is 
\begin{align}
    \sqrt{\frac{1}{n_{\rm samp}}\sum_{j=0}^{n_{\rm samp}-1} z_{k-1-j}^2}.
\end{align}
The EWMA given by \eqref{eq:z_rms_def} thus provides a computationally inexpensive approximation with an effective window length given by \eqref{eq:gamma_nsamp}.

Let $z_{{\rm rms}, {\rm stop}} > 0$ be the user-specified \textit{RMS-based stopping threshold}. 
Then, in the case of RCAC, the update equations for $P_k$ and $\theta_k$ given by \eqref{eq:pk_update} and \eqref{eq:theta_update}, respectively, are modified as
\begin{align}
    P_k, \theta_k
        &
            = \begin{cases}
            \eqref{eq:pk_update}, \eqref{eq:theta_update} & z_{{\rm rms}, k} \ge z_{{\rm rms}, {\rm stop}},\\
            P_{k-1}, \theta_{k-1} & {\rm otherwise}.
            \end{cases} 
\end{align}
Hence, once $z_{{\rm rms}, k}$ falls below the chosen threshold, parameter updates are frozen, effectively transitioning RCAC from an adaptive to a fixed-gain controller.
The stopping criterion thus acts as a \textit{persistence} condition, preventing premature termination of learning due to transient noise-induced reductions in the performance variable.

%
The algorithm for updating RCAC with the RMS-based learning stopping criteria is summarized in Algorithm \ref{alg:RCAC_RMS}.

\begin{algorithm}
    \caption{RCAC update with the RMS-based stopping criterion}
    \label{alg:RCAC_RMS}
    \KwIn{
        Performance variable $z_k,$ 
        previous control input $u_{k-1},$ 
        previous RMS $z_{{\rm rms}, k},$
        forgetting factor $\gamma\in(0,1)$,
        RMS-based stopping threshold $z_{{\rm rms}, {\rm stop}}.$
    }
    \KwOut{Commanded control input $u_{\rmc, k}.$}
    Update $\phi_k$ as shown in \eqref{eq:phik}.\\
    Update $\phi_{\rmf, k}$ and $u_{\rmf, k}$ as shown in \eqref{eq:phifk} and \eqref{eq:ufk}, respectively. \\
    Update $z_{{\rm rms}, k}$ as shown in \eqref{eq:z_k_rms_recursive}. \\
    \eIf{$z_{{\rm rms}, k}\geq z_{{\rm rms}, {\rm stop}}$}
    {
        Update $P_k$ and $\theta_k$ as shown in \eqref{eq:pk_update} and \eqref{eq:theta_update}, respectively.\\
    }{
        Freeze $P_k$ and $\theta_k,$ such that $P_k = P_{k-1}$ and $\theta_k = \theta_{k-1}.$
    }
    Update $u_{\rmc, k} = \phi_k \theta_k,$ as shown in \eqref{eq:controller}. \\
\end{algorithm}

\section{Simulation and Experimental Validation}\label{sec:simulation_experiment}

The objective of this section is to validate the proposed RMS-based stopping criterion in both numerical simulation and physical experiments. 
In particular, transverse vibration at the tip of a cantilever beam is suppressed using a noncollocated actuator configuration. 
The control strategy does not rely on an explicit beam model and instead uses measured data to synthesize the controller in real time.

\begin{figure}
\centering
\includegraphics[width = 0.8\columnwidth]{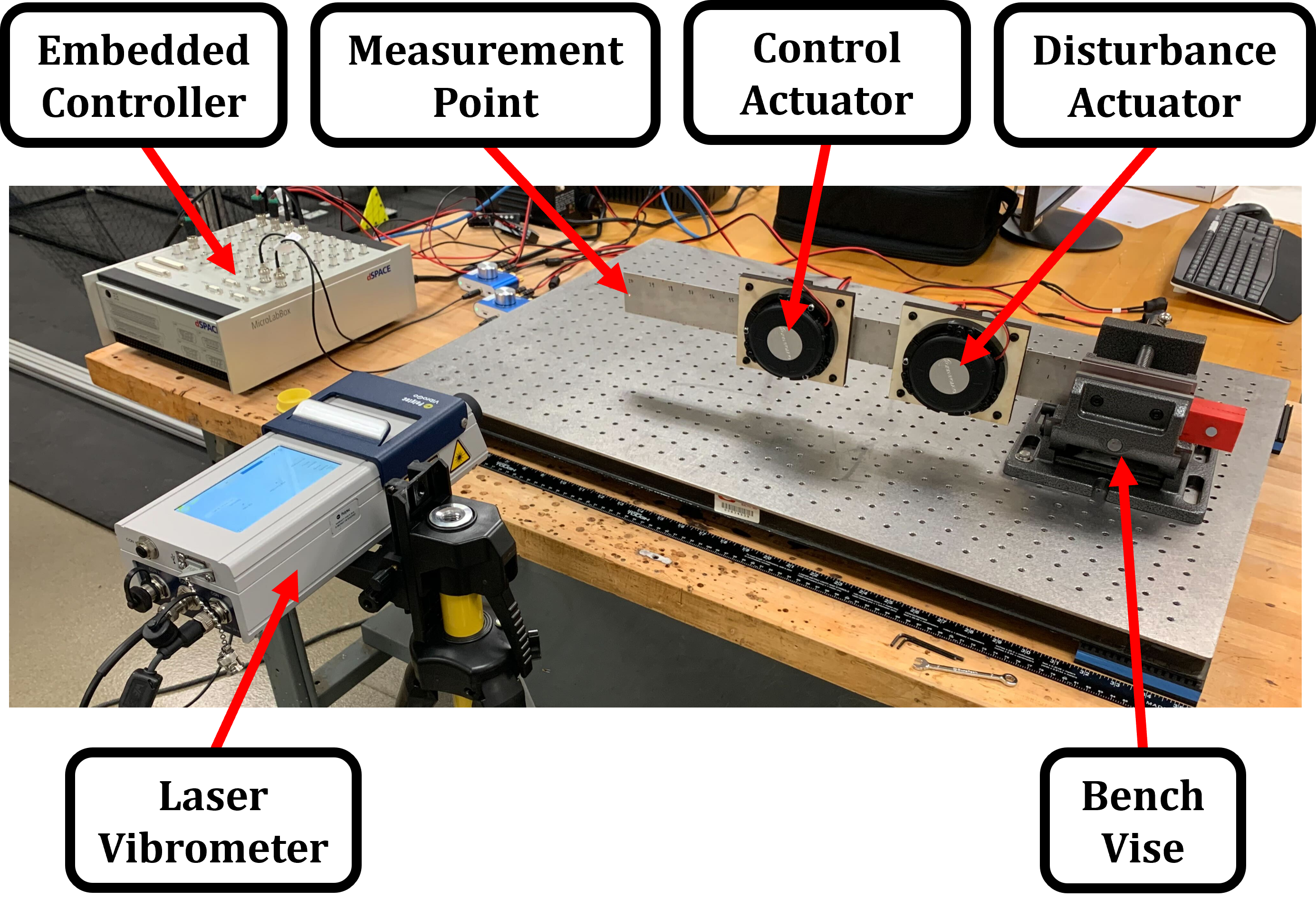}
\caption{Cantilever beam experimental setup consisting of a horizontally mounted aluminum beam clamped by a bench vise, a disturbance shaker near the fixed end, a control shaker near midspan, a laser vibrometer measuring transverse tip displacement, and a dSPACE MicroLabBox II system for real-time control and data acquisition.}
\label{fig:cantilever_experiment}
\end{figure}

The experimental platform consists of a horizontally mounted aluminum cantilever beam, as shown in Figure \ref{fig:cantilever_experiment}. 
The beam has a length of $0.5$ m (measured from the clamped boundary to the free tip) and a rectangular cross-section with height $0.05$ m and thickness $0.001$ m. 
The clamped boundary condition is enforced using a bench vise to realize a fixed support.

Two inertial actuators are used to apply forces to the beam: one serving as the disturbance source and the other as the control actuator. 
Both actuators are 8-$\Omega$ Dayton Audio TT25-8 bass shakers mounted to the beam via a custom-fabricated coupling assembly constructed from laser-cut basswood plates with bolted joints.
Self-clinching nuts are used to prevent loosening during operation. 
The tightening torque is limited to $0.4$ $\rm N\cdot m$ using a calibrated torque meter to ensure repeatable mounting conditions while avoiding excessive preload that could reduce force transmission efficiency.

Transverse beam displacement is measured at the tip using a VibroGo scanning laser vibrometer. 
The measured signal is processed in real time using a dSPACE MicroLabBox II embedded control system, which provides integrated analog-to-digital (A/D) and digital-to-analog (D/A) interfaces. 
The generated control and disturbance commands are amplified by separate 80 W Facmogu F900S power amplifiers before being applied to the actuators.

\begin{figure}
\centering
    \resizebox{\columnwidth}{!}{%
    \begin{tikzpicture}[>={stealth'}, line width = 0.5mm]
        \node [smallblock, inner sep=0.0em, fill=none, line width = 0.75mm,  minimum height = 0.6cm , minimum width = 0.7cm, label = {270:\scriptsize Cantilever Beam}] (system) {\centering \includegraphics[width=12em]{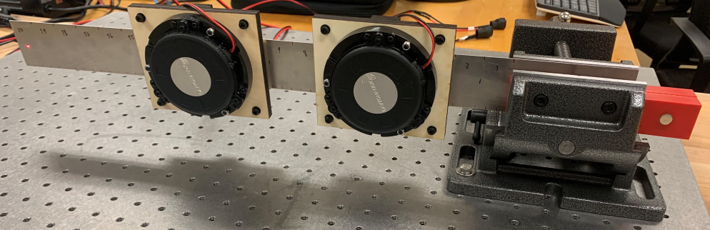}};
        \node [smallblock, inner sep=0.25em, fill=none, line width = 0.75mm, minimum height = 0.6cm , minimum width = 0.7cm, left = 2em of system.west, label = {270:\scriptsize Input Shaker}] (shaker_inp) {\centering \includegraphics[width=3em]{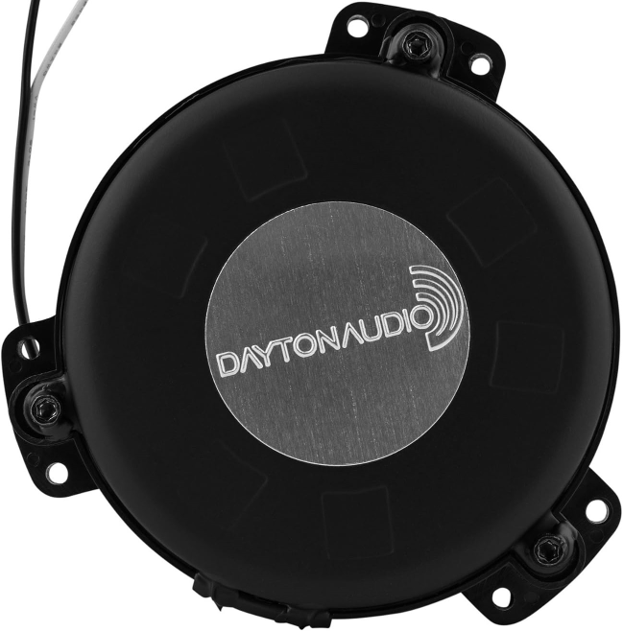}};
        \node [smallblock, inner sep=0.25em, fill=none, line width = 0.75mm, minimum height = 0.6cm , minimum width = 0.7cm, above = 2.5em of shaker_inp.north, label = {270:\scriptsize Disturbance Shaker}] (shaker_dist) {\centering \includegraphics[width=3em]{Figures/Bass_Shaker.png}};
        \node [smallblock, inner sep=0.1em, fill=none, line width = 0.75mm, minimum height = 0.6cm , minimum width = 0.7cm, right = 2em of system.east, label = {270:\scriptsize Laser Vibrometer}] (vibrometer) {\centering \includegraphics[width=5em]{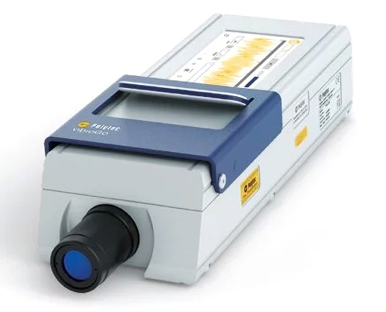}};
        \node [smallblock, inner sep=0.1em, fill=none, line width = 0.75mm, below right = 3em and -1em of system.south, label = {270:\scriptsize Digital Computer}] (controller) {\centering \includegraphics[width=8em]{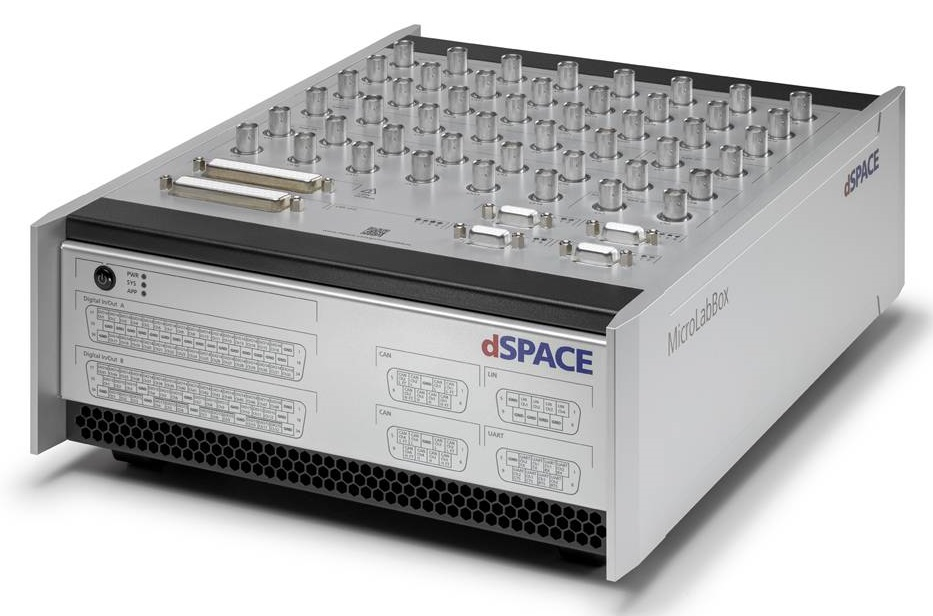}};
         \node [smallblock, inner sep=0.25em, fill=none, line width = 0.75mm, minimum height = 0.6cm , minimum width = 0.7cm, above left = 1em and 4em of controller.west, label = {270:\scriptsize Input Amplifier}] (amp_inp) {\centering \includegraphics[width=3em]{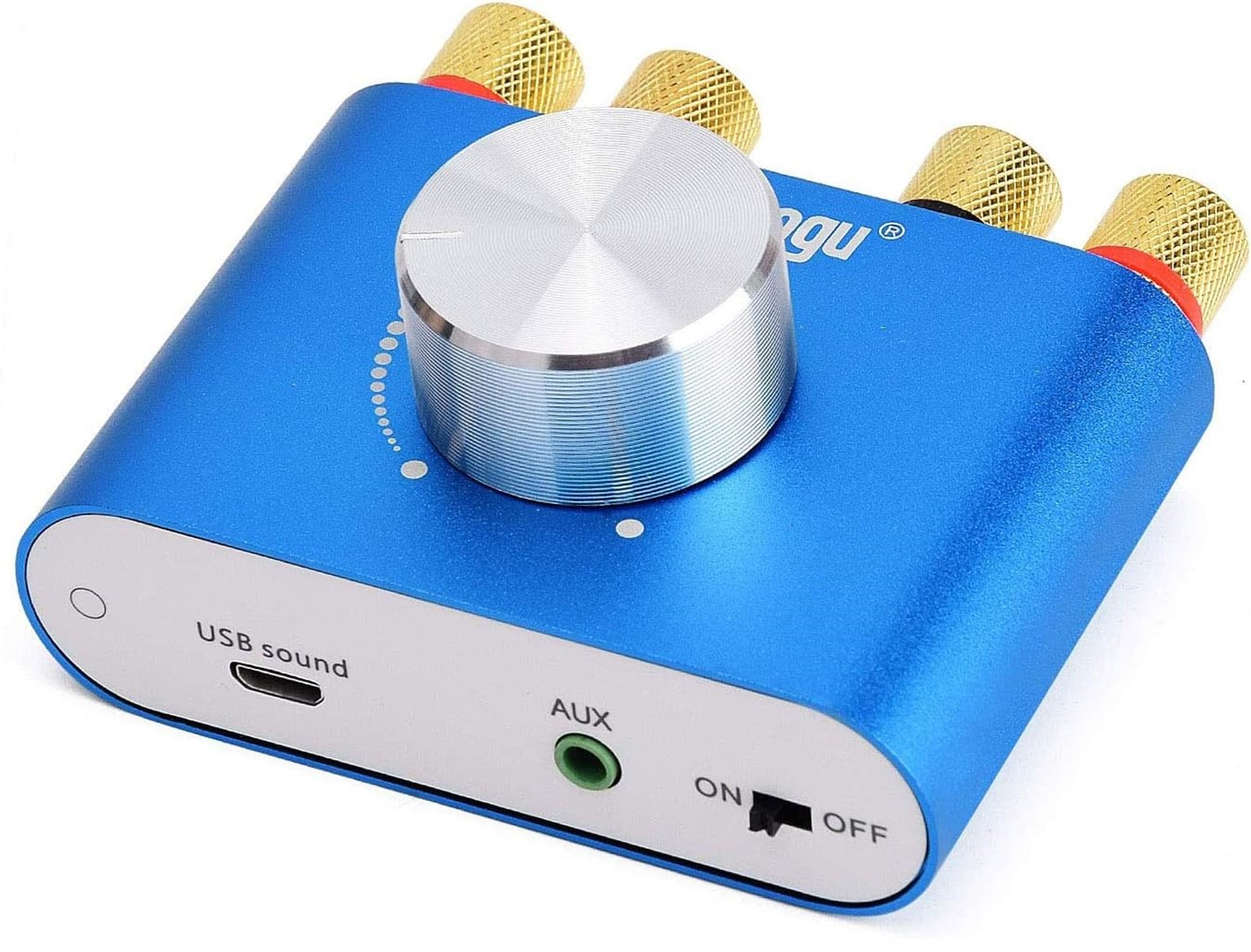}};
        \node [smallblock, inner sep=0.25em, fill=none, line width = 0.75mm, minimum height = 0.6cm , minimum width = 0.7cm, below left = 1em and 4em of controller.west, label = {270:\scriptsize Disturbance Amplifier}] (amp_dist) {\centering \includegraphics[width=3em]{Figures/Amplifier.png}};
        \draw[green!60!black, ->] ([yshift = -1em]controller.west) -| ([xshift = 3em]amp_dist.east) -- node[above, xshift = 0.5em]{$d$} (amp_dist.east);
        \draw[blue!80!black,->] ([yshift = 1em]controller.west) -| ([xshift = 3em]amp_inp.east) -- node[above, xshift = 0.5em]{$u$} (amp_inp.east);
        \draw[blue!80!black,->] (amp_inp.west) -| ([xshift = -1.25em]shaker_inp.west) -- (shaker_inp.west);
        \draw[green!60!black,->] (amp_dist.west) -| ([xshift = -2.5em]shaker_dist.west) -- (shaker_dist.west);
        \draw[blue!80!black,->] (shaker_inp.east) -- (system.west);
        \draw[green!60!black,->] (shaker_dist.east) -| (system.north);
        \draw[red!80!black,->] (system.east) -- (vibrometer.west);
        \draw[red!80!black,->] (vibrometer.east) -- ([xshift = 1em]vibrometer.east) |- node[above, very near end, xshift = -0.25em]{$y$} (controller.east);
    \end{tikzpicture}
    }
    \caption{Block diagram of sensing, control, and actuation hardware. The measured displacement $y$ from the laser vibrometer is processed by the dSPACE MicroLabBox II, which computes the adaptive control signal $u$. Independent power amplifiers drive the control and disturbance shakers mounted on the cantilever beam.}
    \label{fig:cantilever_experiment_hardware}
\end{figure}

A schematic of the hardware interconnections, including sensing, signal conditioning, control hardware, and actuation, is shown in Figure \ref{fig:cantilever_experiment_hardware}.

\subsection{Lumped-Parameter Simulation Model}\label{subsec:LPM}

\begin{figure}
\centering
\includegraphics[width = 0.9\columnwidth]{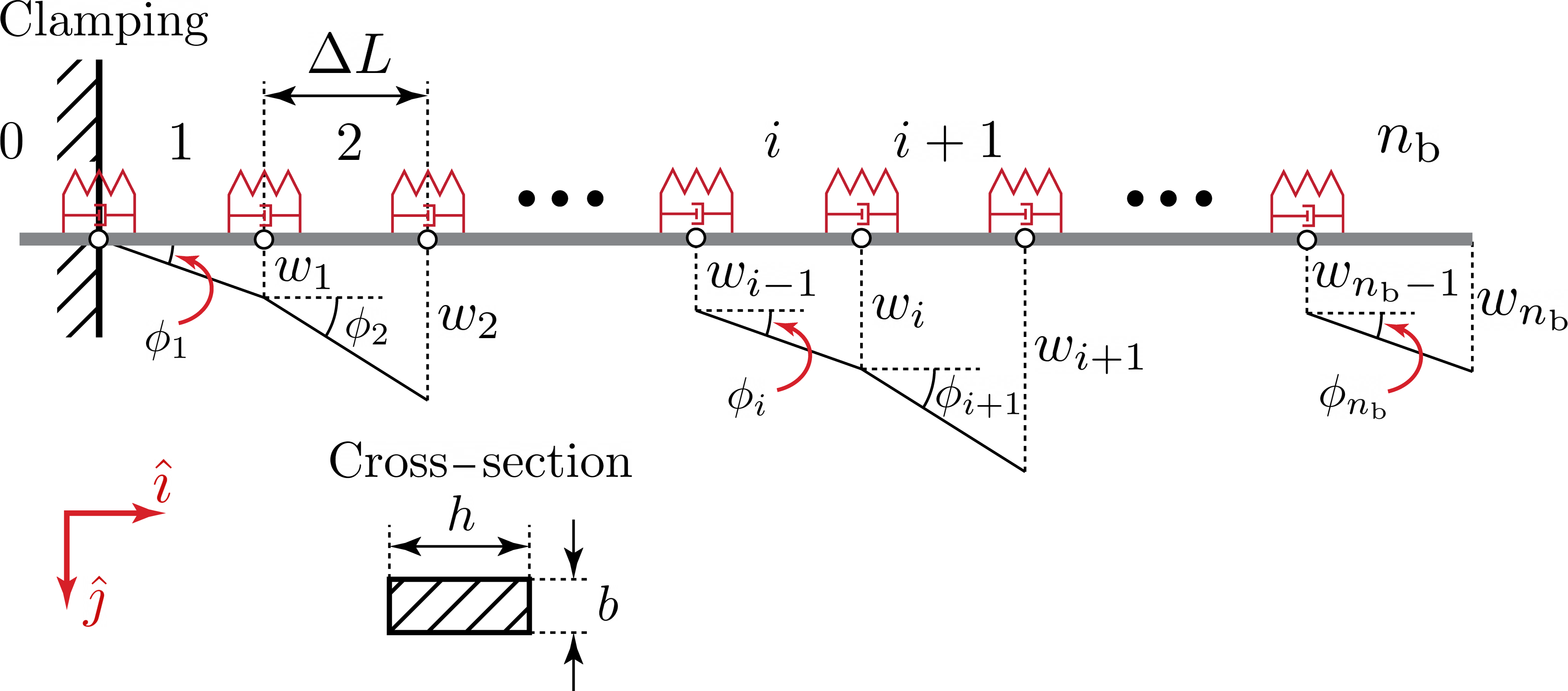}
\caption{Lumped parameter model of a cantilever beam.}
\label{fig:cantilever_LPM}
\end{figure}

The controller is first validated in simulation using a lumped-parameter model (LPM) of the cantilever beam, shown in Figure \ref{fig:cantilever_LPM}. 
The baseline LPM is derived and described in detail in \citep{paredes2026model}. 
In this work, the LPM serves as a preliminary testbed for evaluating the proposed stopping criterion before experimental implementation. 
Although certain model parameters are chosen to reflect the physical cantilever setup, the primary objective of the LPM is not to replicate the experiment exactly, but rather to create conditions under which parameter drift is likely to occur, thereby enabling controlled evaluation of the stopping mechanism.

To better reflect the experimental configuration, the masses of the control and disturbance actuators are explicitly incorporated into the simulation model by augmenting the beam mass matrix with their respective 
contributions. Specifically, the modified mass matrix is defined as
\begin{equation}
    M_{\rm{new}} 
    = 
    M 
    + 
    m_u M_{i_u} 
    + 
    m_d M_{i_d},
\end{equation}
where $M \in \mathbb{R}^{n_\rmb \times n_\rmb}$ is the nominal mass matrix of the $n_\rmb$-element lumped-parameter beam model, $m_u$ and $m_d$ denote the control and disturbance actuator masses, respectively, and $M_{i_u}$ and $M_{i_d}$ represent their spatial mass distributions within the discretized beam.
The actuator mass distribution matrix associated with node $i$ is defined as
\begin{equation*}
    M_i 
    \isdef 
    \frac{1}{2}
    \left(
        e_{n_\rmb,i-1} e_{n_\rmb,i-1}^\rmT
        +
        e_{n_\rmb,i} e_{n_\rmb,i}^\rmT
    \right),
\end{equation*}
where $e_{n_\rmb,i}$ denotes the $i$th column of the $n_\rmb \times n_\rmb$ identity matrix.
Table \ref{tab:LP_param_table} lists the parameters used in the simulation model,
where all parameters correspond to the experimental configuration, except $\alpha$ and $\beta$, which are chosen so that the maximum damping ratio of the eigenvalues corresponding to the system matrix of the LPM is around $6.5 \times 10^{-3}$ to achieve a sufficiently underdamped response suitable for inducing parameter drift and evaluating the stopping criteria.

\begin{table}[ht]
    \centering
    \caption{Lumped-parameter cantilever beam model parameters corresponding to the experimental configuration and used in simulation.}
    \label{tab:LP_param_table}
    \begin{tabularx}{\columnwidth}{p{0.165\columnwidth} X p{0.25\columnwidth}}
    \hline
    \textbf{Parameter}
    & 
    \textbf{Definition}
    &
    \textbf{Value}
    \\
    \hline
        $
            L
        $ 
        & 
        Beam length.
        &
        0.5 m
    \\
        $
            h
        $ 
        & 
        Beam cross-section height.
        &
        0.05 m
    \\
        $
            b
        $ 
        & 
        Beam cross-section width.
        &
        0.001 m
    \\
        $
            m
        $ 
        & 
        Beam mass.
        &
        0.07 kg
    \\
        $
            m_{i_\rmu}, m_{i_\rmd}
        $ 
        & 
        Shaker actuator mass.
        &
        0.43 kg
    \\
        $
            E
        $ 
        & 
        \hspace{-0.875em} $\begin{array}{l} \mbox{Aluminum Young's} \\ \mbox{modulus.} \end{array}$
        &
        69 $\times$ $10^9$ N/m$^2$
    \\
        $
            \alpha
        $ 
        & 
        \hspace{-0.875em} $\begin{array}{l} \mbox{Stiffness-proportional} \\ \mbox{damping coefficient.} \end{array}$
        &
        1.5 $\times$ $10^{-3}$ s$^{-1}$
    \\
        $
            \beta
        $ 
        & 
        \hspace{-0.875em} $\begin{array}{l} \mbox{Mass-proportional} \\ \mbox{damping coefficient.} \end{array}$
        &
        2.5 $\times$ $10^{-7}$ s
    \\
        $
            n_\rmb
        $ 
        & 
        Number of beam elements.
        &
        20
    \\
        $
            i_\rmu
        $ 
        & 
        Input beam element.
        &
        12
    \\
        $
            i_\rmd
        $ 
        &
        \hspace{-0.875em} $\begin{array}{l} \mbox{Disturbance beam} \\ \mbox{element.} \end{array}$
        &
        5
    \\
        $
            i_\rmy
        $ 
        & 
        \hspace{-0.875em} $\begin{array}{l} \mbox{Measurement beam} \\ \mbox{element.} \end{array}$
        &
        20
    \\
    \hline
    \end{tabularx}
\end{table}

Closed-loop numerical simulations are conducted using the LPM with the adaptive controller described in Section \ref{sec:RCAC}. 
The simulations are implemented in the Simulink environment using the \texttt{ode45} solver with relative and absolute tolerances set to $10^{-8}$. 
Appropriate transition blocks are used to interface the continuous-time plant dynamics with the discrete-time controller. 
A sinusoidal disturbance
\begin{align}
    d(t) = \sin(2\pi f_\rmd t),
    \label{eq:dist}
\end{align}
with frequency $f_\rmd = 20$ Hz, is applied near the clamped boundary of the beam. 
The control objective is to attenuate the resulting transverse vibration at the beam tip by modulating the control input $u$ applied near the beam midspan.

Control implementation parameters are specified as follows. 
In the adaptive controller, we set 
$l_\rmc = 15$, 
$p_0 = 5 \times 10^{-3}$, and
the control weighting $R_u = 0.05$.
The sampling period is $T_\rms = 0.005$ s,
and the input saturation is set such that $u_k \in [-4, 4]$ N.

For the target model $G_\rmf$, the parameters are chosen to account for the sinusoidal disturbance, the phase lag introduced by the noncollocated actuator and sensor configuration, and the sign of the LPM leading coefficient. 
Specifically, 
$\omega_\rmf = 2\pi f_\rmd$ rad/s, 
$\alpha_\rmf = 0.9$, 
$d_\rmf = 2$, 
and $N = -1$.

Measurement noise is added to the simulated displacement signal to promote conditions under which parameter drift is more likely to occur, resulting in a signal-to-noise ratio (SNR) of approximately $15$ dB for $z_k$. 
To ensure a smooth RMS estimate in the stopping criterion, the threshold is set to $z_{\rm rms,stop} = 0.05$, and the forgetting factor $\gamma$ is selected such that the effective window length is $n_{\rm samp} = 500$. 
The initial RMS value for recursive computation is set to $z_{{\rm rms},0} = 0,$ corresponding to the absence of prior performance history.

\begin{figure}
\centering
\includegraphics[width = \columnwidth]{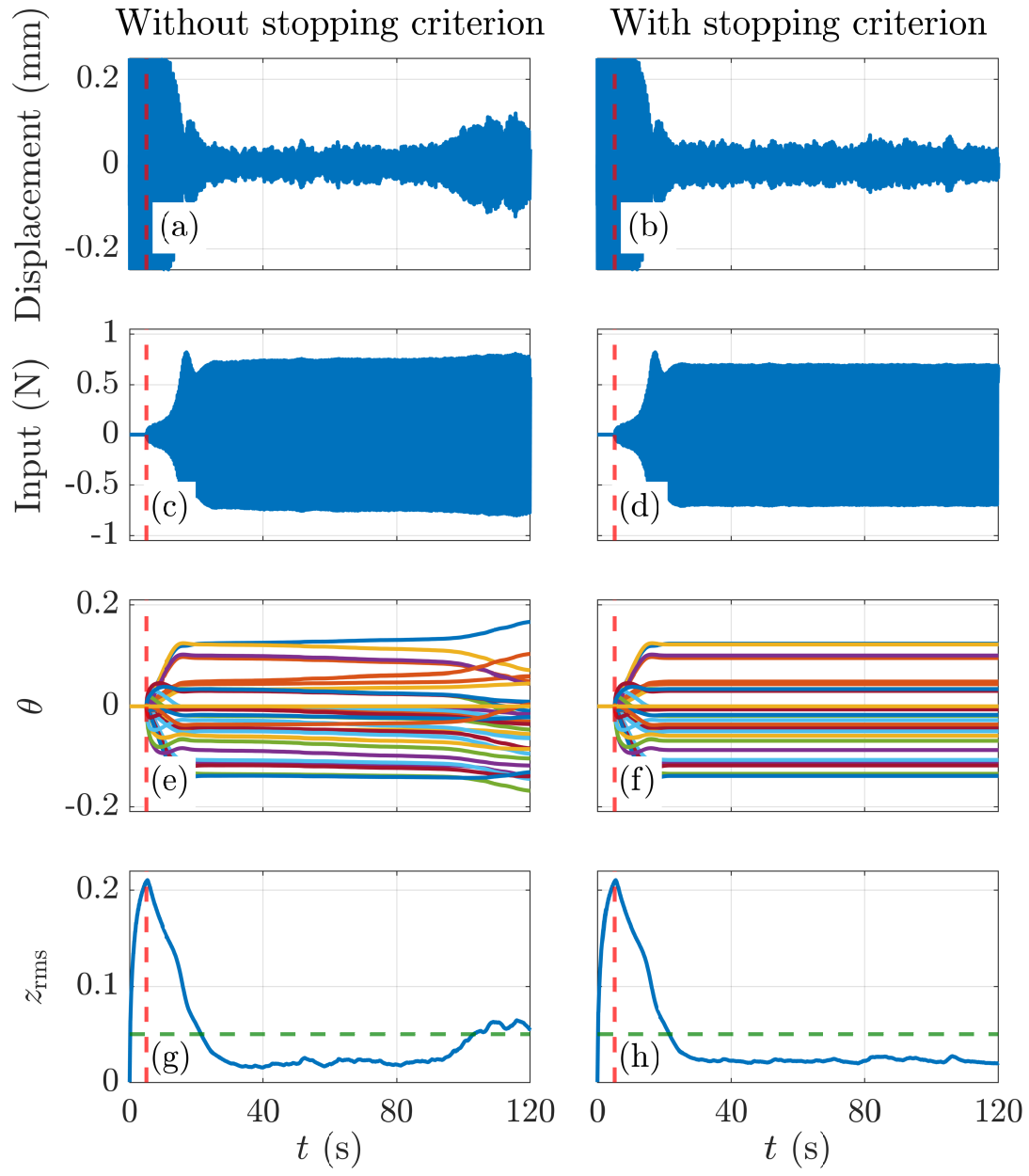}
\caption{Closed-loop numerical simulation results using RCAC
where (a), (c), (e), (g) and (b), (d), (f), (h) correspond to the cases
without and with the RMS-based learning stopping criteria,
respectively.
(a), (b) show the displacement versus time, (c), (d) show the input versus time, (e), (f) show $\theta$ versus time, and (g), (h) show $z_{\rm rms}$ versus time.
The vertical dashed red line indicates the time at which RCAC is enabled.
The horizontal dashed green line in the $z_{\rm rms}$ versus $t$ plots indicates the value of the RMS-based stopping threshold $z_{{\rm rms}, {\rm stop}}.$
}
\label{fig:sim_results}
\end{figure}

\begin{figure}
\centering
\includegraphics[width = \columnwidth]{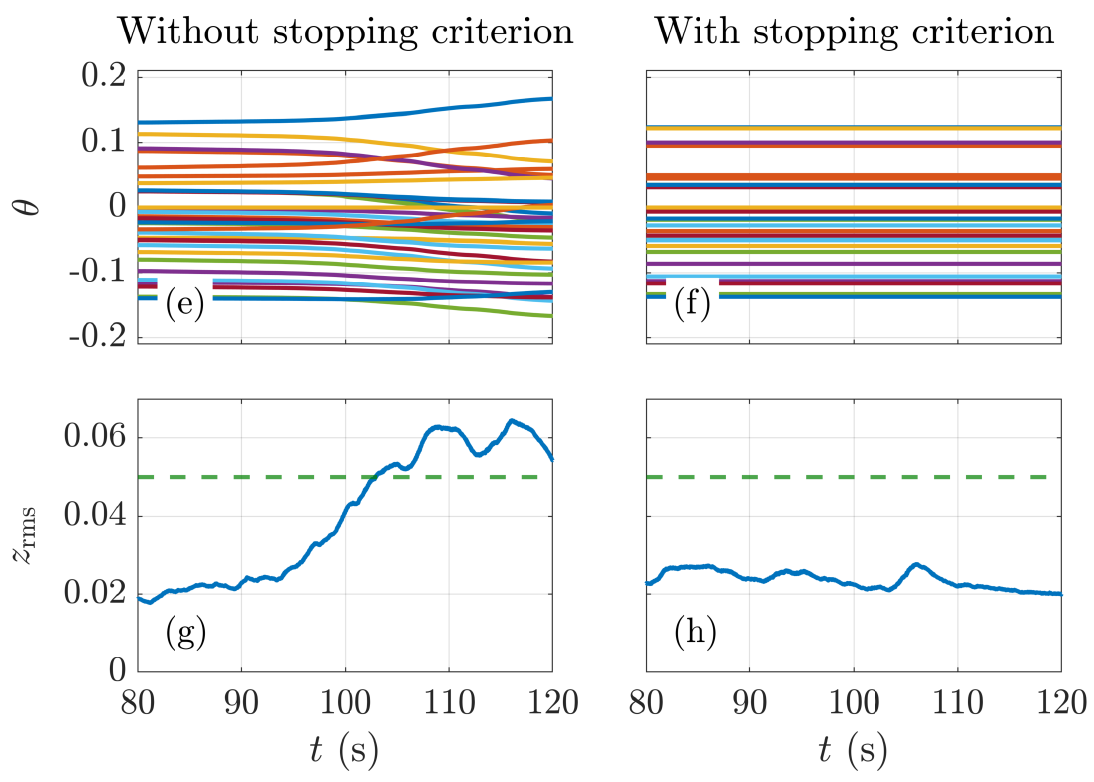}
\caption{Detail of the controller parameter vector $\theta$ and the RMS signal $z_{\rm rms}$ from Figure \ref{fig:sim_results}, showing their evolution near the end of the simulation.
The horizontal dashed green lines denote the RMS-based stopping threshold $z_{{\rm rms},{\rm stop}}$.}
\label{fig:sim_results_theta_40_60_s}
\end{figure}

The controller is activated $5$ seconds after the start of the simulation to allow the disturbance-induced vibrations to reach steady oscillatory behavior.
The RMS calculation is initiated at the start of the simulation, before the controller is activated, to allow $z_{{\rms},k}$ to capture the steady-state amplitude of the disturbance-induced vibrations.
Figure \ref{fig:sim_results} shows the closed-loop simulation results obtained using RCAC with and without the RMS-based learning stopping criterion.
The first row shows the transverse tip displacement, the second row shows the control input applied by the actuator, the third row shows the controller gains adapted by the RCAC algorithm, and the fourth row shows the corresponding RMS value.
To provide a clearer view of the parameter evolution and RMS behavior near the end of the simulation, the results corresponding to the third and fourth rows are shown in greater detail in Figure \ref{fig:sim_results_theta_40_60_s}.

Note that without the RMS-based stopping criterion, parameter adaptation continues even after satisfactory vibration suppression has been achieved, ultimately leading to performance degradation.
In contrast, when the RMS-based stopping criterion is employed, adaptation is halted once the RMS value falls below the prescribed threshold, thereby preventing parameter drift and avoiding the onset of closed-loop performance degradation.

\subsection{Closed-Loop Experimental Results}

Closed-loop experiments are conducted on the cantilever platform using the adaptive controller described in Section \ref{sec:RCAC}.
All signals are generated and acquired using a dSPACE MicroLabBox II embedded system configured with three processor cores operating in parallel. 

The first core generates the disturbance signal \eqref{eq:dist} applied to the disturbance actuator via a zero-order hold with sampling period $5 \times 10^{-4}$ s. 
A second core samples the laser vibrometer displacement measurements at $5 \times 10^{-4}$ s for high-resolution acquisition and visualization. 
The third core implements the RCAC controller with sampling period $T_\rms = 0.005$ s, consistent with the simulation framework. 
Within the controller, the displacement signal is high-pass filtered to remove bias and low-pass filtered to attenuate high-frequency measurement noise.
As in the simulation, the control objective is to attenuate transverse tip vibrations induced by $d$ by modulating the control input $u$ applied near the beam midspan.
Note that, in contrast to the simulation, in which the control and disturbance signals were modeled as forces, in the experimental setup, these signals correspond to the voltages applied to the shakers.

Control implementation parameters are specified as follows. 
In the adaptive controller, we set 
$l_\rmc = 15$, 
$p_0 = 1 $, and
the control weighting $R_u = 0.1$. 
The sampling period is $T_\rms = 0.005$ s,
and the input saturation is set such that $u_k \in [-8, 8]$ V to prevent the input from exceeding the voltage output limits from the embedded system.
The target model $G_\rmf$ and its associated parameters are chosen identically to those used in the simulation study described in Section \ref{subsec:LPM}. 
In the stopping criterion, the threshold is set to $z_{\rm rms,stop} = 0.05$, and the forgetting factor $\gamma$ is selected such that $n_{\rm samp} = 10$, based on preliminary open-loop measurements. 
This value was chosen based on preliminary open-loop experiments conducted with the disturbance applied and the controller disabled. 
The resulting displacement data were used to evaluate different values of $n_{\rm samp}$, and $n_{\rm samp} = 10$ was found to provide sufficient smoothing of variations in $z_{{\rm rms},k}$ without introducing excessive delay in detecting changes in vibration amplitude.
The initial RMS value is set to $z_{{\rm rms},0} = 0.$
The controller is enabled after the disturbance-induced oscillations reach steady-state behavior.
The RMS calculation is initiated before the controller is enabled to allow $z_{{\rms},k}$ to initially capture the steady-state amplitude of the disturbance-induced vibrations.
It is important to emphasize that some controller parameters differ from those used in simulation. 
As discussed in Section \ref{subsec:LPM}, the simulation model is designed to promote conditions under which parameter drift is likely to occur and does not replicate the full experimental dynamics. 
In particular, the LPM excludes actuator dynamics and nonlinear effects introduced by the mechanical mounting assembly.
Accordingly, the LPM and physical experiment are treated as distinct validation platforms, each requiring independently tuned control parameters.

\begin{figure}
\centering
\includegraphics[width = \columnwidth]{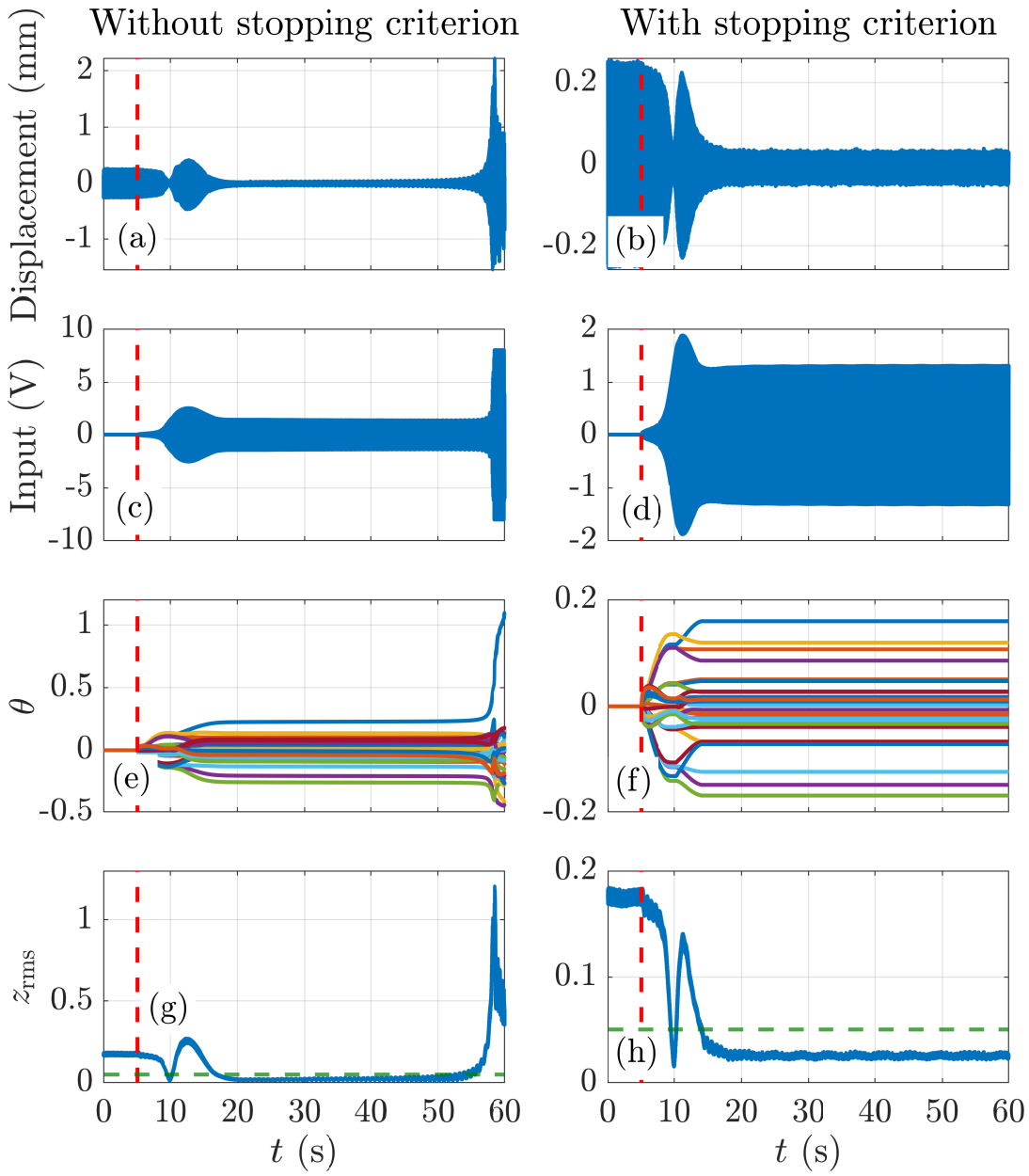}
\caption{Closed-loop experimental results using RCAC
where (a), (c), (e), (g) and (b), (d), (f), (h) correspond to the cases
without and with the RMS-based learning stopping criteria,
respectively.
(a), (b) show the displacement versus time, (c), (d) show the input versus time, (e), (f) show $\theta$ versus time, and (g), (h) show $z_{\rm rms}$ versus time.
The vertical dashed red line indicates the time at which RCAC is enabled.
The horizontal dashed green line in the $z_{\rm rms}$ versus $t$ plots indicates the value of the RMS-based stopping threshold $z_{{\rm rms}, {\rm stop}}.$
Note that the vertical axis scales differ between the two columns.
}
\label{fig:exp_results}
\end{figure}

\begin{figure}
\centering
\includegraphics[width = \columnwidth]{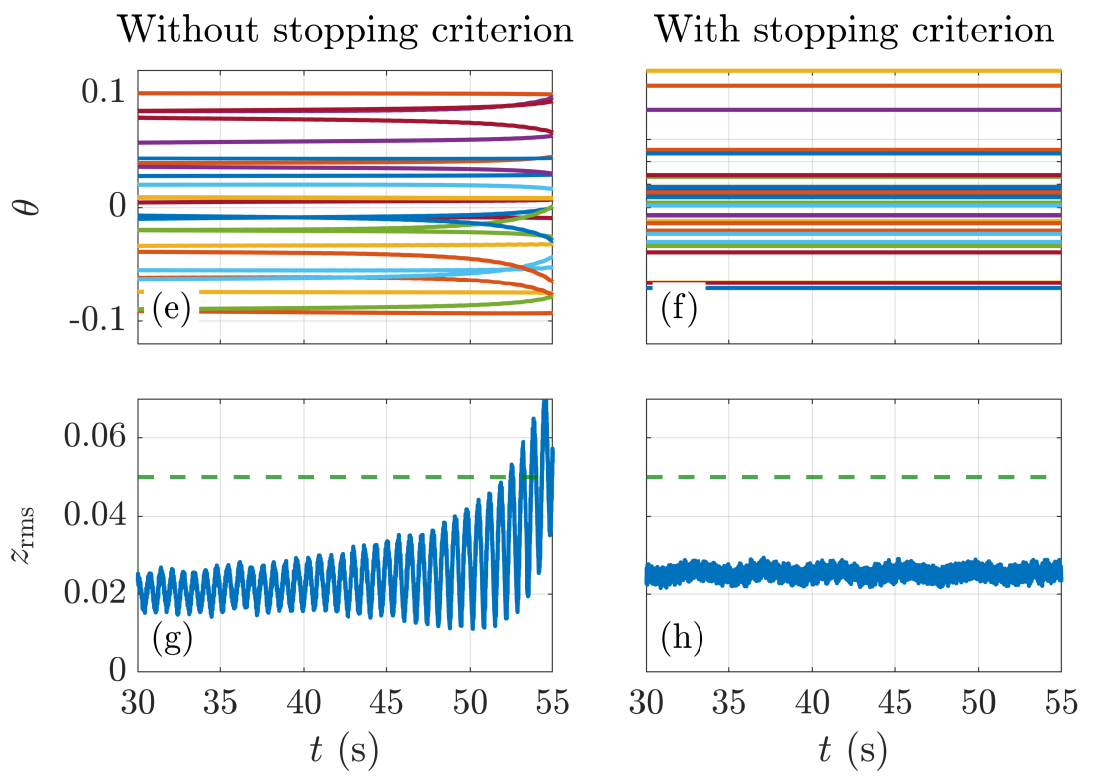}
\caption{Detail of the controller parameter vector $\theta$ and the RMS signal $z_{\rm rms}$ from Figure \ref{fig:exp_results}, showing their evolution near the end of the experiment.
The horizontal dashed green lines denote the RMS-based stopping threshold $z_{{\rm rms},{\rm stop}}$.}
\label{fig:exp_results_theta_30_55_s}
\end{figure}

Figure \ref{fig:exp_results} shows the corresponding closed-loop experimental results, demonstrating that the behavior observed in simulation is replicated on the physical platform.%
The first row shows the measured transverse tip displacement, the second row shows the control input applied by the actuator, the third row shows the controller gains adapted online by the RCAC algorithm, and the fourth row shows the corresponding RMS value computed from the measured response.
A detailed view of the parameter and RMS evolution near the end of the experiment is shown in Figure \ref{fig:exp_results_theta_30_55_s}.

Consistent with the simulation results, when the RMS-based stopping criterion is not used, parameter adaptation continues beyond satisfactory disturbance attenuation, leading to performance degradation.
In contrast, when the RMS-based stopping criterion is enabled, adaptation terminates once the RMS value falls below the prescribed threshold, thereby preventing parameter drift and maintaining stable closed-loop performance in the experimental setting.

These results demonstrate that the RMS-based stopping criterion preserves the performance gains of adaptive control while eliminating parameter drift in both numerical and experimental settings.

\section{Conclusions} \label{sec:conclusions}
This paper presented a computationally efficient criterion, based on a recursively computed RMS error metric, to prevent parameter drift in adaptive control systems.
Unlike traditional approaches that modify the adaptive law through projection operators, normalization, or leakage terms, the proposed method uses a simple performance-based trigger derived directly from measured signals to terminate adaptation once satisfactory vibration attenuation has been sustained over a prescribed interval.
The resulting implementation is lightweight, requires minimal tuning, and is well-suited for real-time embedded applications.

The effectiveness of the approach was demonstrated experimentally on a cantilever beam vibration-suppression problem using a noncollocated actuator configuration.
In the absence of the proposed criterion, sustained excitation conditions led to gradual parameter drift and ultimately degraded performance, and in the worst case, instability of the closed loop. 
By contrast, the modified adaptive algorithm maintained bounded parameter estimates and preserved closed-loop stability without sacrificing transient performance or vibration attenuation capability. 
The experimental results confirm that the proposed criterion prevents the onset of instability while retaining the adaptation benefits necessary for vibration suppression.

An important practical feature of the method is that it does not require an explicit plant model and does not significantly increase the computational burden. 
The RMS metric is computed recursively using the measured signals already available in the controller, making the approach attractive for high-rate control systems with limited computational resources.
Additionally, the method is agnostic to the specific adaptive law and can be incorporated into a wide class of gradient-based or model-reference adaptive controllers.
Furthermore, the use of an RMS metric is particularly well-suited for vibration suppression problems, since vibration responses are inherently oscillatory and typically do not converge to zero even under effective vibration suppression. 
By measuring the energy of the oscillatory response rather than instantaneous error values, the RMS metric provides a natural and robust indicator of sustained vibration attenuation.

Future work will investigate analytical guarantees for boundedness under broader excitation conditions and extension of the approach to multi-mode flexible structures and distributed vibration control systems.

\printcredits

\section*{Declaration of competing interest}

The authors declare that they have no known competing financial interests or personal relationships that could have appeared to influence the work reported in this paper.

\section*{Acknowledgment}

The authors gratefully acknowledge support from the UMBC Technology Catalyst Fund, which provided funding for the procurement of the laser Doppler vibrometer used in the experimental validation of this work.

\bibliographystyle{elsarticle-num}

\bibliography{bib_paper}

\end{document}